\documentclass[12pt,preprint]{aastex}

\begin{document}                                                                     

\title{The Search for Low-mass Companions of B Stars  in the Carina
  Nebula Cluster Trumpler 16
  \altaffilmark{1} }


\author{Nancy Remage Evans }
\affil{Smithsonian Astrophysical Observatory,  MS 4,   
 60 Garden St., Cambridge, MA 02138; email nevans@cfa.harvard.edu}

\author{Kathleen DeGioia-Eastwood }  
\affil{Department of Physics and Astronomy, Northern Arizona
  University, Box 6010, Flagstaff AZ, 86011-6010}

\author{Marc Gagn\'e} 
\affil{Department of Geology and Astronomy, West Chester University,
  West Chester, PA 19883}

\author{Leisa Townsley }  
\affil{Department of Astronomy \& Astrophysics, 525 Davey Laboratory,
  Pennsylvania State University, University Park, PA 16802}

\author{Patrick Broos }  
\affil{Department of Astronomy \& Astrophysics, 525 Davey Laboratory,
  Pennsylvania State University, University Park, PA 16802}

\author{Scott Wolk }  
\affil{Smithsonian Astrophysical Observatory,  MS 70,   
 60 Garden St., Cambridge, MA 02138}

\author{Ya\"el Naz\'e }  
\affil{GAPHE Departement AGO, Universite de Liege, Allee du 6 Aout 17,
Bat. B5C, B4000-Liege, Belgium}

\author{Michael Corcoran }  
\affil{NASA's Goddard Space Flight Center,Code 662, Greenbelt, MD 20771 }

\author{Lida Oskinova }  
\affil{Institute for Physics and Astronomy, University Pots-
dam, 14476 Potsdam, Germany }

\author{Anthony F. J. Moffat }  
\affil{Dept. de Physique, Univ. de Montreal, CP 6128
  Succ. A. Centre-Ville, Montreal, QC H3C 3J7, Canada }

\author{Junfeng Wang }  
\affil{Smithsonian Astrophysical Observatory,  MS 06,   
 60 Garden St., Cambridge, MA 02138}

\author{Nolan R. Walborn }  
\affil{Space Telescope Science Institute   
 3700 San Martin Dr., Baltimore, MD 21218 }

\email{nevans@cfa.harvard.edu}


\altaffiltext{1}{Based on observations made with the {\em Chandra X-ray Observatory} }


\begin{abstract}
We have developed lists of likely B3--A0 stars (called ``late B''
stars) in the young cluster
Trumpler 16.  The following criteria were used: 
location within 3' of $\eta$ Car, an
appropriate V and B-V combination, and proper motion (where
available). Color and magnitude cuts have been made assuming an E(B-V) =
0.55 mag $\pm$ 0.1, which is a good approximation close to the center
of Trumpler 16.  These lists have been cross-correlated with X-ray sources
found in the {\em Chandra} Carina Complex Project (CCCP).  Previous studies have shown
that only very rarely (if at all) do late main sequence
B stars produce X-rays. We present evidence that the
X-ray detected sources are binaries with low-mass companions, since
 stars less massive than 1.4  $M_\odot$ are strong X-ray
sources at the age of the cluster. Both the
median X-ray energies and X-ray luminosities of these sources are in
good agreement with values for typical
 low-mass coronal X-ray sources.
We find that 39\% of the late B stars based on a list 
 with proper motions  have low-mass companions.  Similarly, 
 32\% of a sample without proper motions
 have low-mass companions.   We discuss the X-ray detection
completeness. These results on low-mass companions of intermediate
mass stars are complementary
to spectroscopic and interferometric results, and probe
new parameter space of low mass companions at all separations.  
They do not support a steeply rising
distribution of mass ratios to low masses for 
intermediate-mass  (5$M_\odot$) primaries,
such as would be found by random pairing from the Initial Mass
Function.

\end{abstract}




\section{INTRODUCTION}

B stars later than spectral type B2 do not in general produce X-rays.
In the deepest exposure of a young star cluster 
 (the {\em Chandra Orion Ultradeep Project}, COUP project),
Stelzer, et al. (2005) found  late B and A stars which were not
detected  with log L$_X$ $<$27.6 ergs s$^{-1}$. X-rays in O and B
stars in  the Carina region are discussed 
by Naz\'e, et al. (2011) and Gagn\'e. et
al.  (2011) and references cited therein.  Hotter, more luminous stars
with substantial winds (O and early B stars)
produce X-rays through embedded wind shocks,
colliding winds in binaries, or magnetically confined wind shocks (see
Gagn\'e, et al.). 

Stelzer et al.\ (2005) 
   analyzed the properties of late B (B5 and later) through A stars
in the Orion Nebula Cluster COUP project
(which they call `` weak wind'' stars)
which were detected in X-rays and compared them to
 low-mass T Tau stars.  The luminosity, temperature (relatively 
hard), and variability (flare like) were all consistent with X-rays 
produced by a T Tau companion rather than the B star itself, and Stelzer et al.
concluded that the X-rays are  most probably from late 
spectral type companions. 
On the other hand, in 4 
out of 11 late B and A  stars discussed by Stelzer, et al., 
no X-rays are detected ``several orders 
of magnitude'' below that seen in other weak wind stars (i.e., from 
low-mass companions).   These 4 stars  (36\% of the sample) that 
have no X-rays to a very low level clearly do not 
have a low-mass stellar companion.  
 
The exception to the general lack of X-rays in late B stars appears to
be associated with the rare class of magnetic chemically peculiar
stars.  The model proposed to explain, for instance IQ Aur (a magnetic
A0 star) is the magnetically confined wind-shock model (MCWS, Babel and
Montmerle, 1997).  In this model wind streams from opposite
hemispheres collide at the equator, creating X-rays from strong
shocks.  $\Theta^1$ Ori C is thought to be a high mass analog of this
process (Gagn\'e, et al. 1997).  However, magnetic chemically peculiar
stars are rare and would only be a minor contaminant to the X-ray
detected group of late B stars.  
For instance, Power, et al. (2007) find that magnetic
Ap/Bp stars constitute only 1.7\% of intermediate mass stars within 100
pc of the sun.    
Furthermore, very few late B magnetic chemically peculiar stars have
actually been detected in X-rays (Drake 1998; Leone 1994; Drake, 
Wade, and Linsky 2006).
For instance, Drake et al. (1994) found 3 to 5 detections in about 100 
Bp-Ap stars in the Rosat All Sky Survey.
This is discussed further in Section 6 below.

The aim of the present study is to develop a sample of  intrinsically
X-ray--quiet stars, the late B stars in Trumpler 16 (hereafter Tr
16).  The stars from this list
detected in a Chandra image become the list suspected to have low mass
companions--young and X-ray active T Tau stars.  

  Focusing on a 
cluster (Tr 16) within the rich but complicated Carina complex provides 
advantages to interpretation, namely that the stars are confined to a
much smaller age range ($\simeq$ 3 Myr old) than that covered in the whole
region (though some lower mass stars may be older; see Wolk, et al.,
2011). Townsley, et al. (2011) provide an introduction to the
 Carina complex and the Chandra Carina Complex
Project (CCCP).  
The population of massive stars and the  extinction in Tr 16
 have been well studied, and the range of extinction is much smaller
 than for the entire complex. The full complex also contains a wide
 range of stellar densities or environments, of which Tr 16 is
 one of the denser areas.  This is not to say that Tr 16 can be
 simply characterized as a symmetric cluster.  It is, in fact,  a
 grouping of subclusters (Feigelson, et al., 2011; Wolk, et al. 2011),
 one of which contains $\eta$ Car itself.  Previous studies of Tr 16
 are summarized by Wolk, et al., who focus on the low mass stars.  
 The present study pertains to late B stars in Tr 16;
several other studies in this issue discuss B and O stars in the
entire Carina Complex (Naz\'e, et al., 2011 and Gagn\'e et al. 2011).  
Povich, et al. (2011)  discuss young stellar objects of intermediate mass
identified by mid-infrared excess. The range of ages in the entire
complex is large enough to include the  pre-main sequence stars in
that study, as well as main sequence stars here.

The goal here is to estimate the fraction of B3 to A0 stars 
in  Tr 16 that have low-mass companions using the Chandra X-ray data.
  There are several difficulties in identifying these companions by 
other means.  Radial velocities in this spectral 
range are limited in accuracy because of the broad lines.  This is
particularly a problem in identifying companions of (say) 0.5 $M_\odot$   
for a 5 $M_\odot$ primary (mass ratio q = M$_2$/M$_1$ = 0.1).  Similarly, 
for resolved companions, light from the primary overwhelms light from 
the secondary.  Recently, 
interferometric and adaptive optics approaches as well as 
high resolution satellite images have improved the situation by 
resolving a number of
companions, adding information about  companions in wide orbits.
A number of such surveys of B stars are listed in Sch\"oller, et
al. (2010). 
 However, since there is a large population of low
mass field stars,  a resolved red companion has a high probability of
being a chance  
alignment.   X-ray observations identify young companion stars since 
 T Tau stars can easily be distinguished from much older field stars 
which have very weak X-ray flux.  
This is an important strength of the X-ray approach.  In summary, we 
still know almost nothing of low-mass companions of B 
stars.  X-ray observations provide a new way to identify low mass
companions and hence complete the picture of binary properties.

Massive (M $>$ 8 M$_\odot$) and intermediate-mass (8 $> $ M $>$ 3 M$_\odot$)
stars are typically formed as members 
of groups: rich star clusters, sparse clusters, multiple systems, or 
binary systems. This is  important in the
redistribution of angular momentum necessary in cloud collapse.
Thus, the resulting distributions of angular momentum and mass
(the Initial Mass Function or IMF) are shaped by this formation 
environment.  It is important to determine the observed parameters
of these groupings, even though 
 subsequently  clusters and multiple systems 
can be altered by both  internal and external interactions. 
However, to statistically estimate the multiplicity of stars 
is a non-trivial task. Even more difficult is to put
constraints on binary properties, such as $q$,  the mass 
ratio of the secondary to the primary.  Systems with low mass
companions identified in this study provide new information about
these questions.

The discussion below contains the following parts: 
the development of a sample of mid to late B stars,
the detection of X-ray sources in this list, examination of the X-ray 
sources for
corroborative evidence that they are produced by low-mass coronal
sources, and discussion of the results.

\section{SAMPLE OF STARS }

Massive stars in Tr 16 (O and early B stars) have been extensively
studied, using spectra to derive spectral types (see Massey and
Johnson 1993).  Spectra are not available for late B stars,  
 however  photometry and proper motions do exist.
 We have confined our attention to stars within 3' of
$\eta$ Car to obtain a high proportion of cluster members.


We have developed a sample of Tr~16 late B members from two 
sources.  First Cudworth, et al.\ (1993) have determined 
membership from proper motions.  We created a list of 
stars from their list within 3' of $\eta$ Car 
using the Vizier database.  Although Tr
16 is a complex cluster (see Wolk, et al. 2011), this radius should 
contain most of the cluster members.
We have removed from the list any stars found by Cudworth, et al.
to have a membership probability of less than 0.80 
(Fig. 1a).  This is particularly important in removing foreground
contaminants. To create a list of Zero Age Main Sequence
(ZAMS) stars between B3 V and A0 V, we have used the 
ZAMS from Schmidt-Kaler (1982). 
 Fig. 1a (center line) shows the ZAMS 
 using a distance of 2.3 kpc (Smith 2006), and a mean reddening of 
E(B-V) = 0.55 mag, which corresponds to V - M$_V$ = 13.51 mag.
The two other lines have been shifted by -0.1 and +0.1 mag in 
E(B-V), illustrating the approximate range in E(B-V).
 The range between the lines also means some evolution past the
ZAMS is included. 
Specifically only stars 
with V between 12.41 and 14.81 mag and (B-V) bluer than 
0.62 have been retained.  
 Fig. 1a confirms 
that this range of E(B-V) is reasonable, and 
that the range contains most of the likely B star members. 
Feinstein et al. (1973) provide a graphical summary of the reddening
from massive stars in Tr 16 (their Fig. 9), which shows that this
range of reddenings is appropriate.    
We make no attempt to assign spectral types or temperatures to
individual stars because of limited information about reddening;
instead we have created 
a list of stars which are late B stars in Tr 16.
The final list of 31 sources (called ``Late B Cudworth'' below) is
provided 
in Table 1, including the 
Cudworth, et al.\ designation, the coordinates (J2000), V, B-V,
and the distance from $\eta$ Car (all from Vizier).


We have examined the proper motions for the stars identified
as members in Tr~16  in the recent 
USNO CCD Astrograph Catalog (UCAC3).\footnote[1]
{http://www.usno.navy.mil/USNO/astrometry/optical-IR-prod/ucac}
Although there are differences from the Cudworth, et al.\ values,
the UCAC3 proper motions  are all reasonably small, 
appropriate for stars at the distance
of Tr~16.  


The second approach used the photometry of 
DeGioia-Eastwood, et al.\ (2001).  Again, the original
list of stars within
 3' of $\eta$ Car was obtained from Vizier (Fig. 1b).  
Again, stars brighter or fainter in V  or 
redder in (B-V) than the ZAMS B3--A0 range (using the same color-magnitude
range as for the Cudworth sample) have been culled from 
the list.  Figure 4 in DeGioia-Eastwood, et al.\ confirms
that a selection using a restricted E(B-V) provides good 
separation from a comparison background field.
Table 2 contains the final list 
(called ``Late B Eastwood'' below), including 
the coordinates, V, (B-V), and the distance from $\eta$ Car. 

There is some overlap between our filtered Cudworth and 
Eastwood lists (as there is in 
the author lists), however, there are also some differences. 
The Eastwood  photometry is deeper.  
Although the two lists have many stars in common, 
since the Cudworth, et al.\ stars have proper motion 
evidence of cluster membership in addition to appropriate 
color-magnitude values, we have analyzed the 
lists separately.  

The stars in Figures 1a and 1b that fall far outside the 
selected V- (B-V) range for late B stars 
have magnitudes and colors that are generally
appropriate for a foreground population of low-mass stars.  
Note that after culling  by Cudworth's proper
motion criterion, very few stars fell outside the cluster region
in Fig. 1a.  As an aside, it is by no means impossible that there
may be a few locations of very high obscuration, resulting 
in B stars even more highly reddened than the reasonably 
generous limits in Figures 1a and 1b.  Even if we have omitted 
some obscured stars, it makes no difference to the final result. This
would have only decreased the sample size, not altered the 
fraction of  X-ray detections.









\section{X-RAY SOURCES}

Although Tr~16 is central to the region surveyed in the CCCP
(Townsley, et al.\ 2011), an ACIS observation
of 88.4 ksec already existed in the {\em Chandra} archive. This is a
deeper exposure than the typical exposure in CCCP, and hence the
cluster was not reobserved in CCCP.   
  This observation was originally analyzed by
Albacete-Colombo, et al.\ (2008, called AC below).  The current
analysis is complementary to their discussion in two ways. First, the
data extraction was performed using the ACIS Extract package (AE, Broos, et
al. 2010), which identifies fainter sources, and is also consistent
with other CCCP data (Broos, et al. 2011).  Second, the emphasis of
the current study is on mid to late B stars rather than the low-mass
population. The new treatment of these data  increases the completeness
of detections,
which is  pivotal for our present study 
of X-ray emission in intermediate-mass stars. 

Point source extraction was performed as described in Broos, et
al. (2011).  Source detection in the vicinity of Tr 16 is fully
described by Wolk, et al. (2011).
X-ray sources found from the processing for the whole
project were cross-correlated with the late 
B star lists (Tables 1 and 2) to determine which sources
in this sample produce X-rays.  Fig 2 shows the Tr 16 region of the
ACIS image with the late B stars marked.  The B stars detected are
spread fairly homogeneously through the image.  
The late B stars detected in X-rays are listed in Table 3.  
The X-ray/Cudworth offsets for the 12 sources in Table 3 range from 
0.14" to 0.73", with a mean of 0.38".   The X-ray/Eastwood offsets 
for the 12 sources in Table 3 range from 0.08" to 1.1", with a mean of 0.46".

The X-ray sources are evenly spread through the luminosity and 
temperature ranges in  Figures 3a and 3b.  This is as expected for a random
event (the existence of a binary).
One characteristic we investigate
is  the B3 spectral type commonly used as the cutoff  below 
which hot stars do not produce X-rays through wind shocks as O
stars and early B stars do (see discussion in Gagn\'e, et al., 2011 
and Naz\'e et al., 2011).  If the 
hottest stars in the late B sample did produce their own X-rays, we 
would expect a concentration of X-ray sources at the highest
luminosities (the lowest V magnitudes). 
 Instead X-ray sources are evenly spread through all 
luminosities in Figs. 3a and 3b.  For the Cudworth sample, 
the mean V for the stars
detected in X-rays is  13.31 $\pm$ 0.63 mag, as compared to 13.42
$\pm$ 0.50 mag for the undetected stars, and the mean B-V values 
are  0.41 $\pm$ 0.10 mag
(detected) and 0.39 $\pm$ 0.07 mag (undetected).  For the Eastwood
sample, the mean V values are 13.25 $\pm$  0.61 mag (detected) and
13.52 $\pm$  0.56 (undetected).  For B-V, the means are  0.42 $\pm$
0.10 mag (detected) and 0.43 $\pm$  0.09 mag (undetected). 
For both samples, the mean V of the detected sources
is indistinguishable from the mean V of undetected sample within the
uncertainties.  The same is true for B-V.  

Properties for the detected X-ray sources (Table 3) were
derived as follows.
The net number of detected X-ray events, median
X-ray energy, and flux (columns 4, 5, and 6) are standard outputs 
from the AE package (Broos, et al. 2011) for total energy range
(0.5-8.0 keV).  
To deredden the fluxes we have used
the simple procedure of assuming a constant N$_H$ of 3 x 10$^{21}$ cm$^{-2}$,
which corresponds to E(B-V) = 0.52 mag (well within the range of
E(B-V) =
0.55 $\pm$ 0.10 mag used to define the sample; Seward 2000). 
The information required to
compute individual reddenings is not available, and for the sources
within the center of Tr~16, this is a good approximation. Combining
this reddening with a temperature of log T (K) $\simeq$ 7.5 using a PIMMS
Raymond-Smith model, we
obtained a factor of 1.6 to deredden the observed flux. This is a
typical temperature found by Albacete-Colombo, et al. (2008) for low
mass stars in Tr 16.   The
luminosity is computed from the  dereddened fluxes and a distance
of 2.3 kpc (column 6 in Table 3; Smith 2006).  (If the sources were
actually much softer, typical of massive stars, the absorption
would be larger, resulting in an increase in computed  luminosity of 
about 0.2 in the log, for a logT = 6.9 K.)

\section{SOURCE PROPERTIES}

Is there any corroborating information to support the idea 
that X-ray emission from the sources in Figures 3a and 3b arises from 
low-mass companions rather than from hot star wind shocks?  Figure 4 shows
 the median X-ray energy from the standard extraction process 
as a function of net counts for X-ray detected sources in both the
Cudworth and Eastwood lists. Note that the median energy is for
photons above 0.5 keV, and is not a thermal plasma temperature from a
spectral fit.  To set the context, the energies for 
the O and early B stars (hotter than B3) from the Skiff catalog are 
shown (see Naz\'e, et al.\ 2011).  The Skiff stars are from the entire
Carina project region, not just from Tr~16.  The Tr~16 late B stars (Table
3) are confined to a small region of the plot at high
X-ray energies and modest count rates.  The high median energies 
in particular are a characteristic expected of coronal/magnetic low
mass pre-main sequence
stars which make up the population of companions.  It is 
not surprising, however, that there is some overlap with the Skiff 
O/early B population, since these stars may also occasionally have low
mass companions.  Some of the Skiff stars from the entire Carina
complex may have higher absorption than the Tr 16 stars. This would
result in fewer soft counts in these cases and a higher median
energy.  This makes the distinctly high median energy of late B stars
(as compared with the majority of the Skiff) stars even more notable.

A study of B stars from the whole CCCP region (Gagn\`e et al.\ 2011)
also finds that many X-ray detections have the characteristics of low
mass coronal sources.  Their Fig. 8 shows a bimodal distribution of
X-ray source fluxes.  Since they include early B stars, a number of
objects in that figure are intrinsic X-ray sources.  However the
sample is dominated by lower X-ray flux sources interpreted as low
mass companions.  The sources in our late B sample (Table 3)
have fluxes appropriate to the lower flux ``companion'' portion of 
Gagn\`e, et al. Fig. 8.  

The X-ray luminosities in Table 3 are also consistent with
those of low-mass stars in Tr~16 (Wolk, et al.\ 2011, 
Fig. 7).  They overlap
with the low luminosity range of the early B/O stars (Naz\'e, et
al. 2011) but there is a clear break in the trend of X-ray luminosity
as a function of bolometric luminosity (Naz\'e, et al., 2011 Fig. 3),
consistent with a different X-ray production mechanism.   

We have examined plots of log Lx as a function of V and also B-V for
both Cudworth and Eastwood datesets.  No relation is seen in any of
the plots, consistent with the proposition that the X-ray and optical 
photons are produced by different stars (optical: B stars, X-rays:
companions).

\subsection{Strong Sources}

For the four strongest X-ray sources, we present the light curves and
spectral fits, which can be examined to see whether they show the
characteristics of low-mass stars.   These have been generated as
standard products of the AE package.

\noindent
{\bf Spectra}

The ``best model'' spectra have been obtained from the CCCP
database and are shown in Fig. 5 for (top to bottom)  FA 69, FA 52, Y 188,
 and FB 238 (Broos, et al.\ 2011; Broos, et
al. 2010). The spectra have been fit with 
single temperature APEC thermal plasma models
 and foreground absorption as described in Broos, et al. (2011).   
Since late B
stars do not have strong winds, no additional circumstellar absorption
is expected.  It is a confirmation of the fitting process that the
N$_H$ in the fits corresponds to the range of E(B-V) expected for the
foreground reddening (0.44 to 0.79 mag). The temperatures are 
kT $\simeq$ 2.43 keV (FA 69), kT $\simeq$ 2.44 keV (FA 52), kT$\simeq$
2.13 keV (Y 188),
 and kT $\simeq$ 2.38 keV (FB 238).  All four 
temperatures are higher than those typically found in massive stars (see,
for instance, Fig. 1 in Naz\'e, et al., 2011).

\noindent
{\bf Light Curves}

Light curves are also shown (Fig. 5) for FA 69, FA 52, Y 188,
 and FB 238.  Flux variability is quantified by a 
p-value\footnote{In statistical 
hypothesis testing, the p-value is the probability of
obtaining a test statistic at least as extreme as the one that was
actually observed 
when the null hypothesis is true.}
for the no-variability hypothesis, estimated via the
Kolmogorov-Smirnov (KS) statistic, 
shown as $P_{KS}$ in the right panels.
Of the four,  FA 52 shows
the strongest evidence of flaring 
 in the light curve or in the median X-ray energy, with weaker evidence
 in FA 69.  
This is roughly consistent with the flare duty cycle from the long COUP 
observation of the Orion Nebula Cluster for both solar mass stars and 
for lower mass stars (Caramazza, et al.\ 2007). 

\subsection{Upper Limits}

As expected, many of the late B star were not detected in the Chandra
image. Since the data come from a single image and are reasonably
close to the center (i.e. the psf is approximately constant), the
upper limits to undetected stars have a small range. 
The source extraction (Broos, et al. 2011) provides a 1
$\sigma$ upper limit to the counts from 0.5 to 8.0 keV.   Values range
from 8.9 to 1.2 counts, with 4 counts as a typical value. This
corresponds to a log L$_X$ of 29.65 erg s$^{-1}$.
This is lower than
 the X-ray luminosities of the detected late B
stars in Table 3.  Furthermore, Gagn\'e et al. (2011) provide
comparisons between detected early B stars and other massive stars 
(their Fig. 5) and early B stars, early B upper limits, and low mass stars
(their Fig. 8).  The fact that the upper limits in 
these figures are lower than the the L$_X$ of the bulk of
the B stars but similar to those of the low mass stars is  consistent with
the undetected stars having a low mass companion (or no companion), the
same as we find here.

\section{RESULTS}

In the Cudworth sample, 39\% of the late B stars were detected in the
{\em Chandra} data (Tables 1 and 3); for the Eastwood sample, 32\% were
detected (Tables 2 and 3). These are the stars that are expected to
have low-mass X-ray active companions.

What fraction of late type companions would be detected in the {\em Chandra}
image? This has been discussed by AC (2008). From
comparison with the COUP very deep exposure of the comparable age Orion Nebula
cluster (Preibisch and Feigelson 2005),  they estimate that detections
are 55\% complete for 0.9 to 1.2 $M_\odot$ stars (which will 
be G stars on the main sequence), 40\%
complete for 0.5 to 0.9 M $M_\odot$ stars (which will become 
main sequence K stars), and only 5\% complete
for less massive stars (to become main sequence M stars).  
For our sample, if we use 5 $M_\odot$ 
as typical of late B stars, this means that we will identify
essentially no M star companions (q = M$_2$/M$_1$ $<$ 0.1), but
approximately half the companions more massive than this but cooler
than mid-F spectral types (roughly to q = 0.3).
 
There are two reasons that our companion detection may actually be
higher than this.  The source detection technique used here
(Broos, et al, 2011; Wolk, et al.\ 2011)
 identified 70\% more sources than
AC.  Fig. 3 in Wolk, et al.\ 2011, (a plot of X-ray flux vs J magnitude)
 shows that sources are detected at least 2
magnitudes fainter than in AC.  This means that we detect stars that will
be M stars on the main sequence.  Thus, stars less massive than 0.5  $M_\odot$
are detected, corresponding to q = 0.1.
The second reason is also illustrated in the same figure in Wolk et
   al.  The F$_X$ vs J relation has a clear lower bound in X-ray
   fluxes  to at least J
   = 16 mag which corresponds to M = 0.8 $M_\odot$ from the Siess
tracks (Wolk, et al., Fig 4).  This indicates that source 
detection is complete in this
   range. A second result from this figure discussed in Wolk, et
   al. pertains to stars in the late B through A spectral range
(J between 12 and 14 mag, which would include our detections in Table
 3).   These stars are {\it not} the fainter X-ray sources. 
That is, they are not the lowest-mass  J = 16 mag coronal sources in
that figure.   Although many low mass stars are detected with J $>$
15 mag,  the late B--A star range (J 12-14 mag) is more sparsely 
populated  for
X-ray flux $<$ -6 (log photons sec$^{-1}$ cm$^{-2}$. 
  The implication is
 that (assuming the X-rays in this group are from low-mass companions)
 the   companions are not the least massive pre-main
 sequence stars in the cluster.  As discussed above, low mass
 companions {\it would be} detected to at least M = 0.8 $M_\odot$, however
 they are lacking in the range J = 12-14 mag. 
 That is, binaries among late B and A stars with very 
small q values are deficient.  
 (Preibisch and Feigelson [2005] and Telleschi et
al. [2007] provide comparable diagrams linking X-ray flux and mass.)

\section{DISCUSSION}

To summarize, 39\%  and 32\%  of the late B stars in the Cudworth
sample (Tables 1 and 3) and
Eastwood  sample  (Tables 2 and 3)  were detected respectively in the
{\em Chandra} data. It is expected that these 
overwhelmingly have low-mass companions. Note that system
{\it multiplicity} could be higher if a low-mass
companion is itself a binary.  

\noindent
{\bf Biases}

What biases are likely to be found in these data?  The most important
is contained in the sample selection.  If a cluster member were omitted
from Tables 1 or 2, for example because of an unusually large
reddening, the result is not affected.  The
sample size is decreased, but it is not contaminated by nonmembers.  For
this reason, the Cudworth sample which has the additional proper
motion criterion is taken to be the more authoritative result,
although it is a valuable confirmation that the two lists provide very
similar results (in part because they have a significant overlap).  

One aspect of the sample and detection process, the detection
completeness, was discussed in the previous section. We conclude that
X-ray detection is likely to be complete through 0.8 $M_\odot$
companions.

Finally, is it possible that some of the X-ray detected stars 
are in fact  MCWS stars which produce X-rays intrinsically.  This
would mean we have over estimated the fraction of low mass companions. 
The discussion of fluxes of B stars (Fig. 8 in Gagn\'e, et al., 2011) 
illustrates that this is not a serious distortion. Although 
the Gagn\'e sample contains
hotter B stars than our sample, they are stars of the same age
analyzed in the same way, making them a good comparison.  Gagn\'e et
al. conclude that the sample is made up of two populations.  The
majority of detected B stars nicely match the distribution in X-ray photon
flux expected if
the X-rays are produced by a low mass companion. However, there is 
in addition a small
group of 14 stars with higher fluxes, which are good candidates for
instrinsic X-ray production, for which the MCWS mechanism is the leading
hypothesis.  If we accept this as the population of intrinsic X-ray
producers in the list of 127 early B stars, the fraction is only 11\%.
Applying this fraction to the late B sample, we would 
expect only 1 of the late B X-ray stars to be
an intrinsic X-ray source and wrongly attributed as having a 
low mass companion.

\noindent
{\bf Binary Fractions}

As an example for comparison, the recent discussion by Mason, et
al. (2009) combines speckle interferometry observations of O and B
stars with spectroscopic binary results.  They find, for instance, 
a binary fraction of 66\% among cluster O stars, taken to be the
sample least altered from  the initial condition.  Our approach 
using X-ray identification cannot, of
course, be used for O stars, since they produce X-rays themselves.
The low-mass companions of B stars identified through X-rays 
in this project would only be
present very infrequently in lists of spectroscopic binaries or
speckle interferometry because of small mass ratios and large magnitude
differences respectively.  Hence our binary fraction is 
complementary and approximately
additive to the Mason, et al.\ result.

A second comparison comes from an International Ultraviolet
Explorer satellite (IUE) survey of 76 Cepheids brighter than 8$^{th}$
mag from 2000 to 3200 \AA\/ (Evans 1992).  Any companion hotter than mid
A spectral type would have been detected.  21\% of the sample had hot
companions. (A statistical correction using stars with known orbital
velocities provided a fraction of 34\%.)  This target list is more
similar in mass to late B stars
than the O star list.  Again, the fraction of companions in 
the photometric survey
of hot star companions is approximately additive to the fraction of
low-mass companions in the present study.   


A full synthesis of the results of the present X-ray companion
detection study and other binary survey techniques is premature and
beyond the scope of this paper.  As discussed above, the Mason, et
al. and Evans studies detect relatively high mass companions while the
present X-ray study detects low mass companions. Since there is little
overlap, the respective
fractions can be simply summed for an approximate total fraction.
This results in a very high fraction of binaries, approaching unity. 
 This is in agreement with, for instance, the discussion of 
Kouwenhoven, et al. (2007) for intermediate mass stars in the Sco OB2
association.
Adaptive Optics (AO) surveys are more difficult to combine with the
X-ray results, since AO surveys reveal companions of all masses but
are not sensitive to close binaries which are typically detected in
spectroscopic surveys.  Furthermore, they may include chance
projections of low mass field stars.  In contrast, X-ray detected
companions are only stars young enough to be physical companions.
 The X-ray approach  detects
companions at all separations but only low mass companions (from 
$~$1.4  $M_\odot$ to 0.8  $M_\odot$). In other
words, there is some overlap in detections with AO surveys but in a 
complicated way.  A number
of AO surveys of B stars have recently been done, as summarized in
Sch\"oller, et al. (2010) typically finding a fraction of
approximately 30\%.  Again, the present X-ray results imply that the
full binary fraction including close binaries is considerably higher
than this.  

The result that 39\% of late B stars have low-mass companions is 
a lower limit to the fraction of low mass companions because of the
limit to X-ray sensitivity.  However,  discussion  of the low-mass
stars in the previous section 
 suggests that the X-ray observations may in
fact have uncovered the majority of companions.  
As mentioned in the
previous section this is an intriguing hint that the low-mass companions 
may favor the more massive stars among the low-mass coronal sources.  
The fraction of low-mass companions is 
smaller than          the fraction of companions produced from random
pairing from the IMF, which rises very 
steeply at low-masses. 

The homogeneous distribution of X-ray sources throughout the late B star range
(Figs. 3a and 3b) supports the interpretation that the X-rays are
produced by a low-mass coronal source rather than a continuation of
the wind shock mechanism in more massive stars.  

In summary, we have used {\em Chandra} X-ray data to identify intermediate
mass  (5 $M_\odot$) 
stars with low-mass companions in Tr~16.  This approach together with
AO surveys are exploring  new
parameter space for reasonably massive binary systems.








\acknowledgments

It is a pleasure to acknowledge many interesting conversations
 from the {\em Chandra} Carina
Large Project teams, particularly the Penn State group and the massive
star group.  We also thank an anonymous referee for suggestions which
improved the presentation of the paper.
 Tables 1 and 2 were generated using data from the CDS VizieR
 interface. NRE and SJW acknowledge support from the {\em Chandra} 
X-ray Center NASA Contract NAS8-03060.
This work is supported by {\em Chandra X-ray 
Observatory} grant GO8-9131X (PI:  L.\ Townsley) 
and by the ACIS Instrument Team contract SV4-74018 (PI:  G.\ Garmire),
issued by the {\em Chandra} X-ray Center, which is operated by the
Smithsonian Astrophysical Observatory for and on behalf of NASA under
contract NAS8-03060.  
AFJM is grateful for financial support from NSERC (Canada) and FQRNT (Quebec).
YN acknowledge support from the Fonds National
de la Recherche Scientifique (Belgium),
the PRODEX XMM and Integral contracts, and
the œôòøAction de Recherche Concert´eeœôòù (CFWBAcademie Wallonie Europe).  
NRW acknowledges support from STScI which is
operated by AURA, Inc., under NASA contract NAS5-26555.

\clearpage

\begin{deluxetable}{lrrrcc}
\tabletypesize{\scriptsize}
\tablecaption{Late B Stars (Cudworth) }
\tablewidth{0pt}
\tablehead{
\colhead{Name} &   \colhead{RA} & \colhead{Dec}  & \colhead{V} &
\colhead{ B-V} & Distance to $\eta$ Car \\
\colhead{} &   \colhead{(h  m  s)} & \colhead{($^o$ '  '')}  & \colhead{(mag)} &
\colhead{(mag)} & \colhead{(')}  \\
}

\startdata
    FB204 &  10 44 42.15 & -59 41 40.3 &  13.89 &   0.36 &   2.71  \\
    FA50 &  10 44 44.37 & -59 42 33.8 &  12.91 &   0.32 &   2.77  \\
    FA47 &  10 44 51.66 & -59 43 14.1 &  12.90 &   0.34 &   2.55 \\
    Y127 &  10 44 53.97 & -59 40 19.2 &  14.34 &   0.56 &   1.43 \\
   FB224 &  10 44 55.13 & -59 42 24.9 &  13.27 &   0.49 &   1.63 \\
    FA41 &  10 44 56.70 & -59 40 24.2 &  12.48 &   0.37 &   1.10 \\
    FA40 &  10 44 56.79 & -59 40  03.0 &  13.30 &   0.44 &   1.36 \\
    FA39 &  10 44 57.97 & -59 40  00.5 &  12.83 &   0.45 &   1.31 \\
    Y213 &  10 44 58.57 & -59 43 33.9 &  14.13 &   0.29 &   2.49 \\
   FB200 &  10 44 58.67 & -59 41 16.1 &  13.50 &   0.32 &   0.58 \\
    FA75 &  10 45  05.03 & -59 42  08.0 &  14.20 &   0.58 &   1.02 \\
    FA68 &  10 45 05.75 & -59 41 24.4 &  12.50 &   0.26 &   0.42 \\
    FA69 &  10 45 07.93 & -59 41 34.6 &  13.03 &   0.41 &   0.74 \\
    FA51 &  10 45  07.97 & -59 39  01.4 &  12.89 &   0.30 &   2.20 \\
    FA52 &  10 45  08.41 & -59 38 47.9 &  12.68 &   0.35 &   2.43 \\
    FA70 &  10 45  09.33 & -59 41 28.8 &  13.33 &   0.37 &   0.84 \\
    Y207 &  10 45 10.06 & -59 43 32.5 &  14.01 &   0.35 &   2.55 \\
    Y188 &  10 45 11.24 & -59 42 34.4 &  13.84 &   0.55 &   1.75 \\
    Y116 &  10 45 12.76 & -59 39  06.8 &  12.90 &   0.37 &   2.36 \\
    Y200 &  10 45 13.30 & -59 42 58.7 &  13.47 &   0.45 &   2.24 \\
    Y206 &  10 45 13.59 & -59 43 32.4 &  14.13 &   0.45 &   2.73 \\
    Y166 &  10 45 14.07 & -59 41 42.4 &  13.07 &   0.38 &   1.48 \\
   FB238 &  10 45 16.25 & -59 41 41.7 &  13.57 &   0.44 &   1.74 \\
   FB239 &  10 45 17.34 & -59 41 20.8 &  13.60 &   0.39 &   1.80 \\
   FB240 &  10 45 18.02 & -59 41 10.0 &  13.82 &   0.41 &   1.87 \\
    Y189 &  10 45 18.99 & -59 42 18.9 &  12.57 &   0.31 &   2.31 \\
    Y164 &  10 45 19.03 & -59 41 43.2 &  14.01 &   0.35 &   2.08 \\
    Y190 &  10 45 20.12 & -59 42  08.9 &  13.97 &   0.45 &   2.36 \\
    Y191 &  10 45 20.61 & -59 42 21.9 &  13.30 &   0.36 &   2.51  \\
    Y163 &  10 45 21.12 & -59 41 44.9 &  12.92 &   0.43 &   2.34 \\
    Y193 &  10 45 24.20 & -59 42 31.9 &  13.40 &   0.41 &   2.99 \\

 \enddata
 
\end{deluxetable}                                                                                  

\clearpage

\begin{deluxetable}{llrrrcl}
\tabletypesize{\scriptsize}
\tablecaption{Late B Stars (Eastwood) }
\tablewidth{0pt}
\tablehead{\colhead{\#} & 
\colhead{RA} & \colhead{Dec}  & \colhead{V} &
\colhead{ B-V} & Distance to $\eta$ Car & \colhead{ Other}  \\
\colhead{ } & \colhead{(h  m  s)} & \colhead{($^o$ '  '')}  & \colhead{(mag)} &
\colhead{(mag)} & \colhead{(')}& \colhead{ID*}   \\
}

\startdata

1 &  10 45 05.75 & -59 41 24.2 &  12.41 &   0.25 &   0.43  & FA68 \\
2 &  10 44 58.65 & -59 41 16.0 &  13.44 &   0.33 &   0.65 & FB200 \\
3 &  10 45  04.43 & -59 41 47.5 &  13.52 &   0.42 &   0.73 & CTr16\_2572 \\
4 &  10 45  07.94 & -59 41 34.2 &  12.96 &   0.42 &   0.74 & FA69 \\
5 &  10 45  08.96 & -59 40 41.0 &  12.87 &   0.26 &   0.78 &  \\
6 &  10 45  09.33 & -59 41 28.5 &  13.28 &   0.38 &   0.83 & FA70  \\
7 &  10 45  00.20 & -59 40  05.9 &  13.25 &   0.32 &   1.06 & CTr16\_1611 \\
8 &  10 44 54.77 & -59 41 24.1 &  13.34 &   0.35 &   1.16 &  \\
9 &  10 45  06.88 & -59 42 16.6 &  14.71 &   0.51 &   1.28 &  \\
10 &  10 44 57.90 & -59 40  00.9 &  12.67 &   0.44 &   1.28 & FA39  \\
11 &  10 44 56.73 & -59 40  03.3 &  13.18 &   0.39 &   1.34 & FA40 \\
12 &  10 44 58.45 & -59 39 48.7 &  12.86 &   0.47 &   1.42 &  \\
13 &  10 44 53.92 & -59 40 19.6 &  14.30 &   0.55 &   1.43 & Y127 \\
14 &  10 44 54.36 & -59 40  01.4 &  14.16 &   0.48 &   1.57 &  \\
15 &  10 44 55.16 & -59 42 24.4 &  13.23 &   0.49 &   1.71 & FB224 \\
16 &  10 45 16.31 & -59 41 41.4 &  13.51 &   0.47 &   1.72 & FB238 \\
17 &  10 45 17.38 & -59 41 20.5 &  13.49 &   0.57 &   1.76 & FB239 \\
18 &  10 45 11.30 & -59 42 33.7 &  13.83 &   0.59 &   1.78 & Y188 \\
19 &  10 45 18.05 & -59 41  09.6 &  13.81 &   0.42 &   1.83 & FB240 \\
20 &  10 45 19.06 & -59 41 42.8 &  14.02 &   0.40 &   2.05 & Y164 \\
21 &  10 45  07.99 & -59 39  02.4 &  12.87 &   0.41 &   2.11 & FA51 \\
22 &  10 44 48.86 & -59 42  08.5 &  13.86 &   0.41 &   2.14  & \\
23 &  10 45 13.35 & -59 42 58.3 &  13.44 &   0.54 &   2.26 & Y200 \\
24 &  10 45 12.77 & -59 39  06.7 &  12.88 &   0.48 &   2.28 & Y116 \\
25  & 10 45  03.92 & -59 43 21.4 &  13.45 &   0.57 &   2.29 &  \\
26 &  10 45 19.02 & -59 42 18.6 &  12.53 &   0.28 &   2.31 & Y189 \\
27 &  10 45 21.15 & -59 41 44.5 &  12.86 &   0.47 &   2.31 & Y163 \\
28 &  10 45  08.44 & -59 38 48.3 &  12.64 &   0.49 &   2.35 & FA52 \\
29 &  10 45 20.18 & -59 42  08.5 &  14.01 &   0.43 &   2.35 & Y190 \\
30 &  10 45 20.64 & -59 42 21.5 &  13.26 &   0.35 &   2.51 & Y191  \\
31 &  10 45 10.12 & -59 43 32.1 &  14.09 &   0.33 &   2.60 & Y207 \\
32  & 10 44 51.68 & -59 43 13.3 &  12.84 &   0.35 &   2.62 & FA47 \\
33 &  10 45 24.81 & -59 40 54.4 &  14.57 &   0.35 &   2.68 &  \\
34 &  10 44 47.16 & -59 39 20.4 &  13.51 &   0.61 &   2.70 &  \\
35 &  10 45 13.62 & -59 43 31.9 &  14.21 &   0.46 &   2.77 & Y206 \\
36  &  10 44 42.16 & -59 41 39.7 &  13.85 &   0.42 &   2.77 & FB204 \\
37 &  10 45 24.23 & -59 42 31.5 &  13.35 &   0.43 &   2.98 & Y193 \\

 \enddata
 
* Cudworth designations, except for CTr16 numbers which are from the
Carina X-ray source list

\end{deluxetable}

\clearpage

\begin{deluxetable}{lllrcrr}
\tabletypesize{\scriptsize}
\tablecaption{X-Ray Detections }
\tablewidth{0pt}
\tablehead{
\colhead{CCCP} & \colhead{List*} & \colhead{Name }  &  \colhead{Net Counts}  &
\colhead{Median X-ray Energy} &
\colhead{Flux} & {Log Lum} \\
\colhead{ID } & \colhead{ }& \colhead{ } & \colhead{  }  & \colhead{(K)} &
\colhead{(erg cm$^{-2}$ s$^{-1}$)} & \colhead{(erg s$^{-1}$)}  \\
}

\startdata

CTr16\_1102 & C & FA50 &  67 &  1.4 &   8.62e-15 &  30.94 \\
CTr16\_1504 & C,E &  FA39 &  38 &  1.3 &   2.87e-15 &  30.46 \\ 
 CTr16\_2669 & C &   FA75 &  41 &  1.3 &   4.54e-15 &  30.66 \\
CTr16\_2770  &C,E &   FA68 &  13 &  1.5 &   2.10e-15 &  30.33 \\
CTr16\_3062  &C,E  &  FA69 & 176 &  1.7 &   2.08e-14 &  31.32 \\
CTr16\_3117  & C,E &   FA52 & 117 &  1.5 &   1.56e-14 &  31.20 \\
CTr16\_3230  & C,E &  Y188 & 100 &  1.5 &   1.15e-14 &  31.07 \\
CTr16\_3288   &C,E  &  Y200 &  14 &  1.9 &   4.23e-15 &  30.63  \\
 CTr16\_3296  &C,E &  Y206 &  25 &  1.9 &   4.69e-15 &  30.68 \\
CTr16\_3334  &C,E &  FB238 & 152 &  1.5 &   1.79e-14 &  31.26 \\
CTr16\_3377  & C,E &  Y189 &  47 &  1.5 &   5.68e-15 &  30.76  \\
 CTr16\_3378   &C,E &  Y164 &  25 &  1.6 &   3.52e-15 &  30.55 \\
CTr16-1611 &   E & & 53 &  1.5 &   6.19e-15 &  30.80  \\
CTr16-2572 &  E & & 11 & 1.5  &   1.21e-15 &  30.09 \\

 \enddata

* C: Cudworth list;  E: Eastwood list

\end{deluxetable}                                                                                  




 

\clearpage

\figcaption[]{(left) The Cudworth sample within 3' of
$\eta$ Car with a membership probability $\ge$ 0.80.
The lines are the ZAMS from Schmidt-Kaler from 
B3 V through A0 V.  The central line is for E(B-V) = 0.55 mag;
lines to the left and right have been shifted by -0.1 and +0.1 
mag, respectively in (B-V).  (V and B-V are in magnitudes in all figures.)
(right) The Eastwood sample  within 3' of
  $\eta$ Car.  
Again, the lines are the ZAMS from Schmidt-Kaler from 
B3 V through A0 V for E(B-V) = 0.45, 0.55, and 0.65 mags 
from left to right. 
   \label{fig1}}

\figcaption[]{The center of Chandra ACIS image with the late B stars marked.  
The image is event data color coded for energy (red = 0.5-2 keV, 
green = 2-7 keV) and smoothed with a gaussian. Thus, 
X-ray sources  appear red if they  
are soft, green if they are hard, yellow for in-between.
 $\eta$ Car is the object in
the middle, which is piled up. The green nearly- 
horizontal line is the readout streak from  $\eta$ Car.  The surrounding red  
diffuse object is the $\eta$ Car X-ray nebula.    Image  
coordinates are  J2000. The  yellow circle outlines the 3' search region.   
  Blue symbols are X-ray detected late B stars (crosses: Cudworth,
Table 3; diamonds: Eastwood, Table 3); purple symbols are late B stars
which were not detected (squares: Cudworth; circles: Eastwood).
   \label{fig2}}

\figcaption[]{(left) The Cudworth sample after cuts in V and B-V. Lines are
  the Schmidt-Kaler ZAMS.  Dots are detected in X-rays; x's are not.
(right) The Eastwood sample after cuts in V and B-V. Lines are
  the Schmidt-Kaler ZAMS.  Dots are detected in X-rays; x's are not. 
   \label{fig3}}                                   

\figcaption[]{Median X-ray energy as a function of log of net counts for
  X-ray sources from ACIS Extract (AE). Dots are sources from both the Cudworth
and Eastwood  lists. x's are sources from the Skiff catalog of O 
and early B stars.    
   \label{fig4}}

\figcaption[]{X-ray spectra and time series for FA69, FA52, Y188, and
  FB238 (top to bottom).
Single-temperature thermal plasma models are overplotted on the 
spectra (left panels).  For the fits N$_H$ is 10$^{22}$ cm $^{-2}$ and
kT is in keV.
Temporal variation in photon flux is depicted by both binned and
adaptively smoothed 
light curves (right panels, solid, left ordinate axis).
Flux variability is quantified by a p-value (see text)
for the no-variability hypothesis, estimated via the
Kolmogorov-Smirnov (KS) statistic, 
shown as $P_{KS}$ in the right panels.
Temporal variation in median X-ray energy is depicted by binned time
series (right panels, 
dotted, right ordinate axis).
Both time series are not corrected for background.
 \label{fig5}}




\plottwo{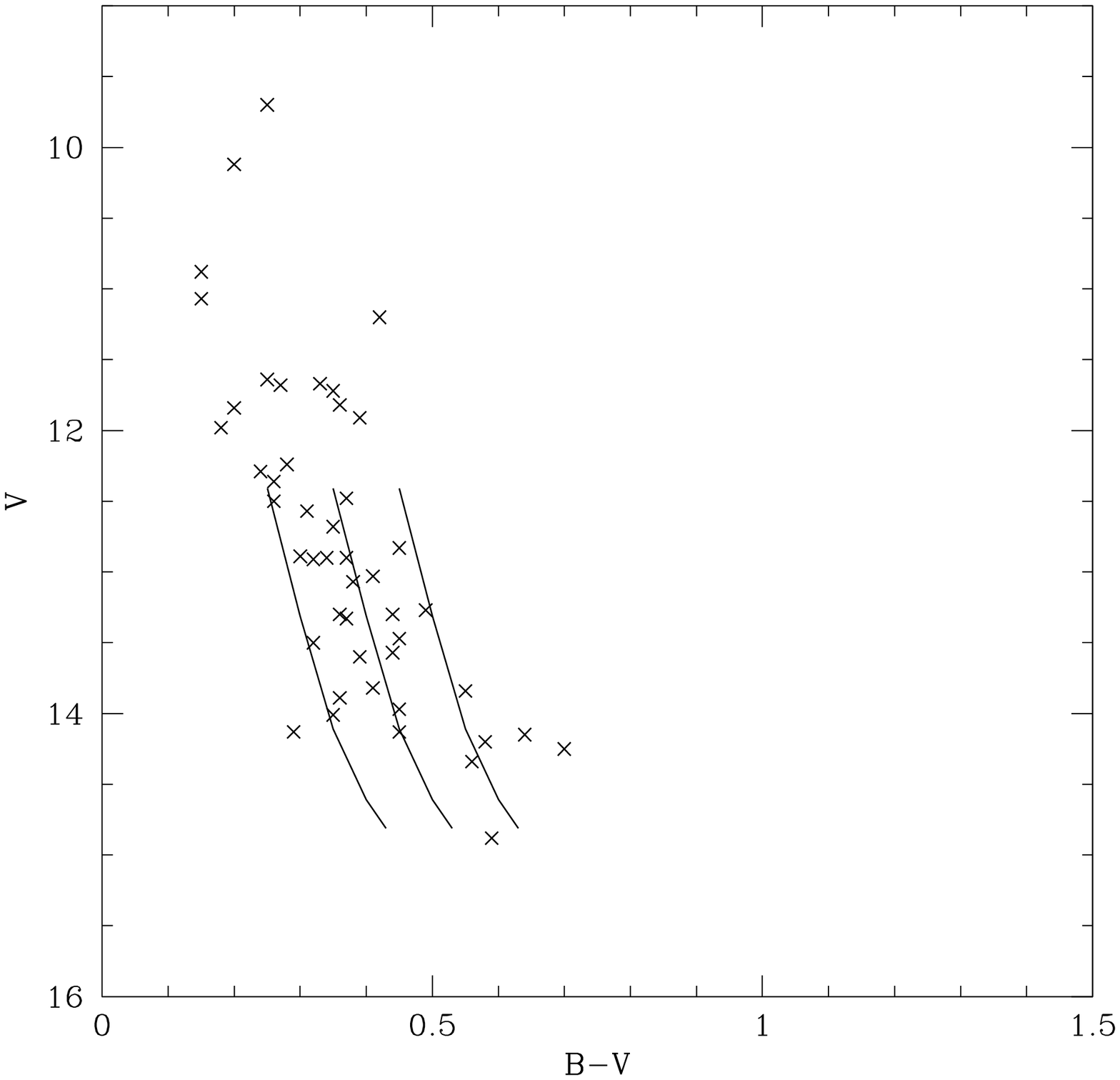}{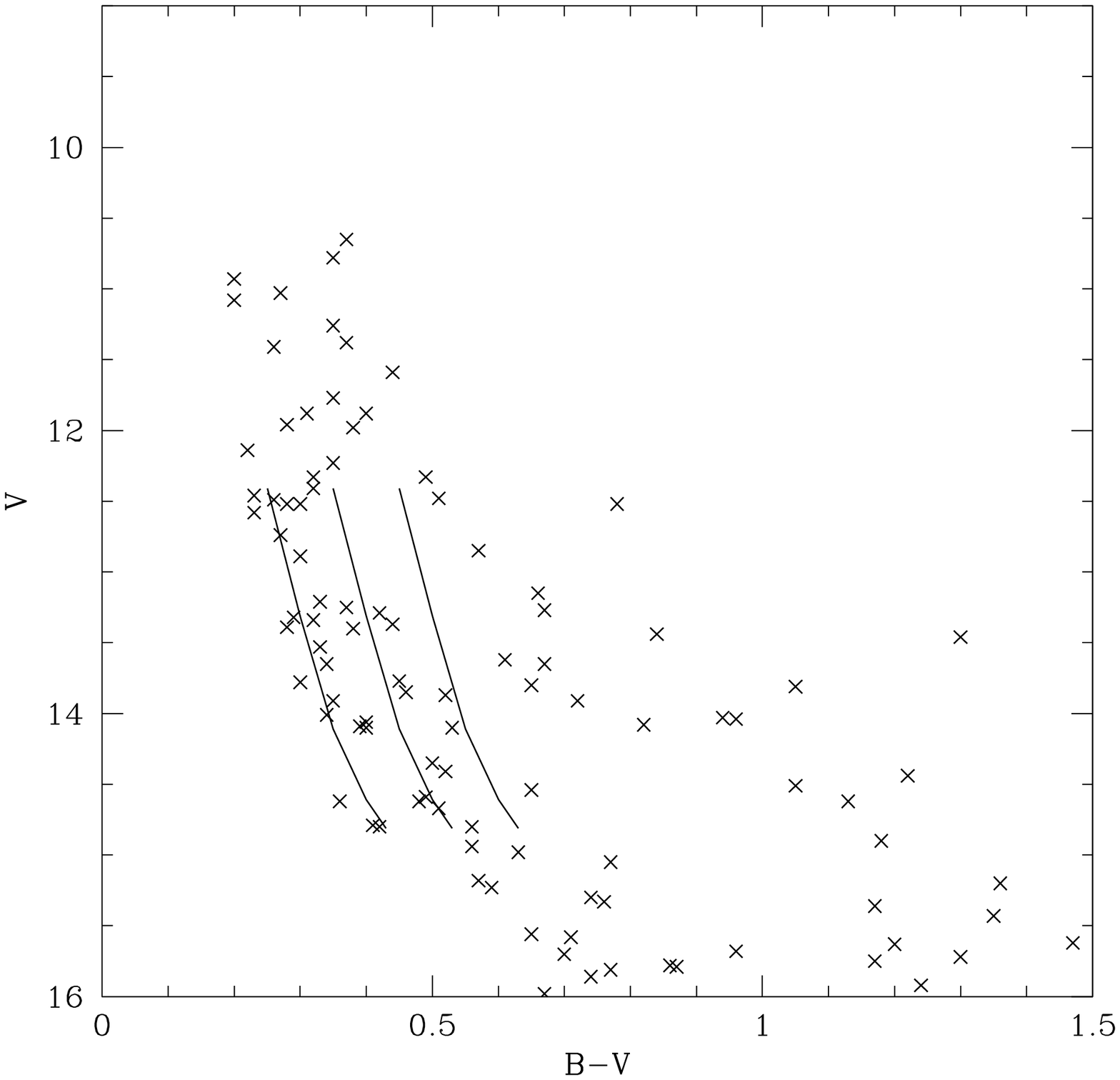}



 

\plotone{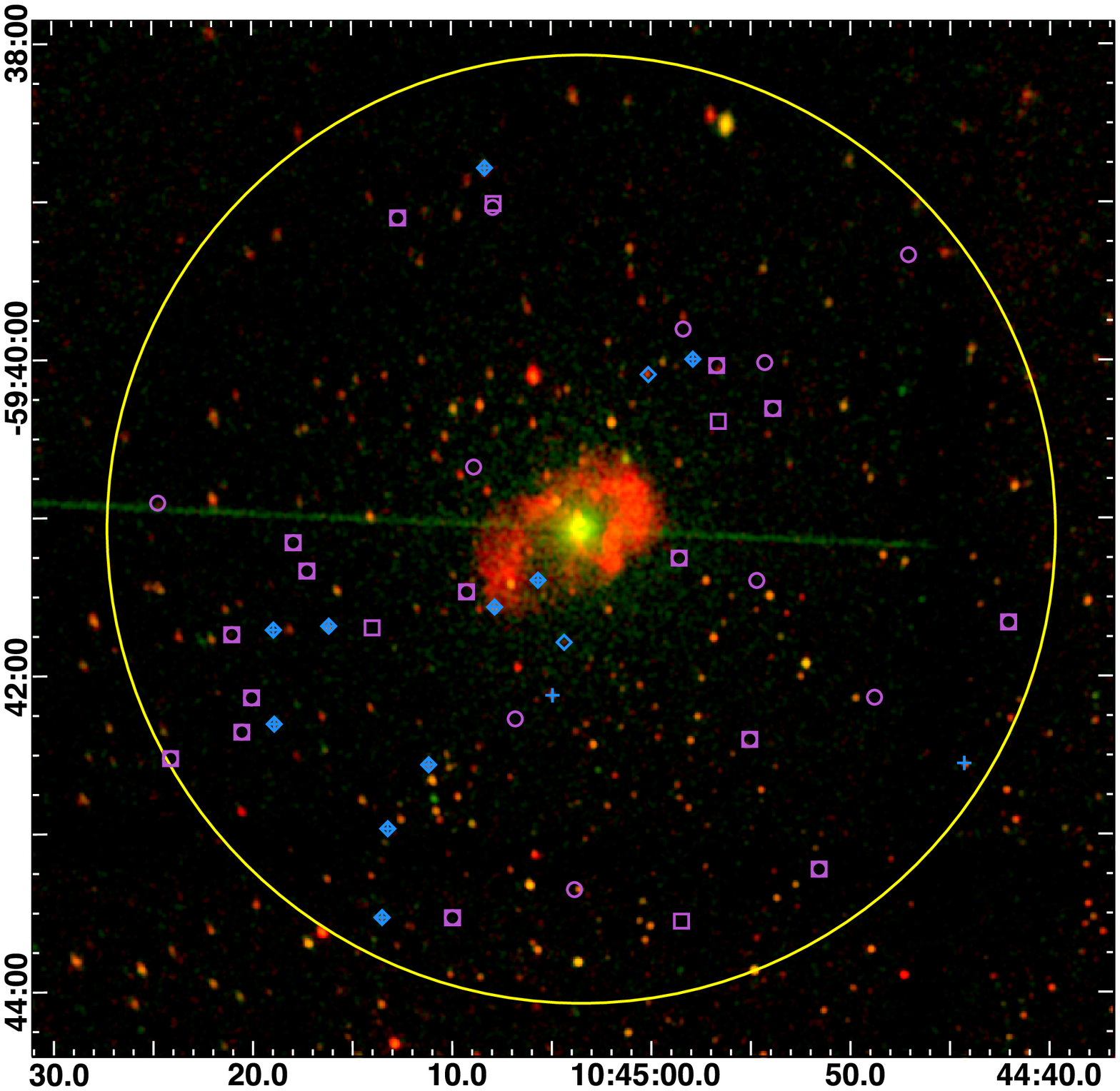}

\plottwo{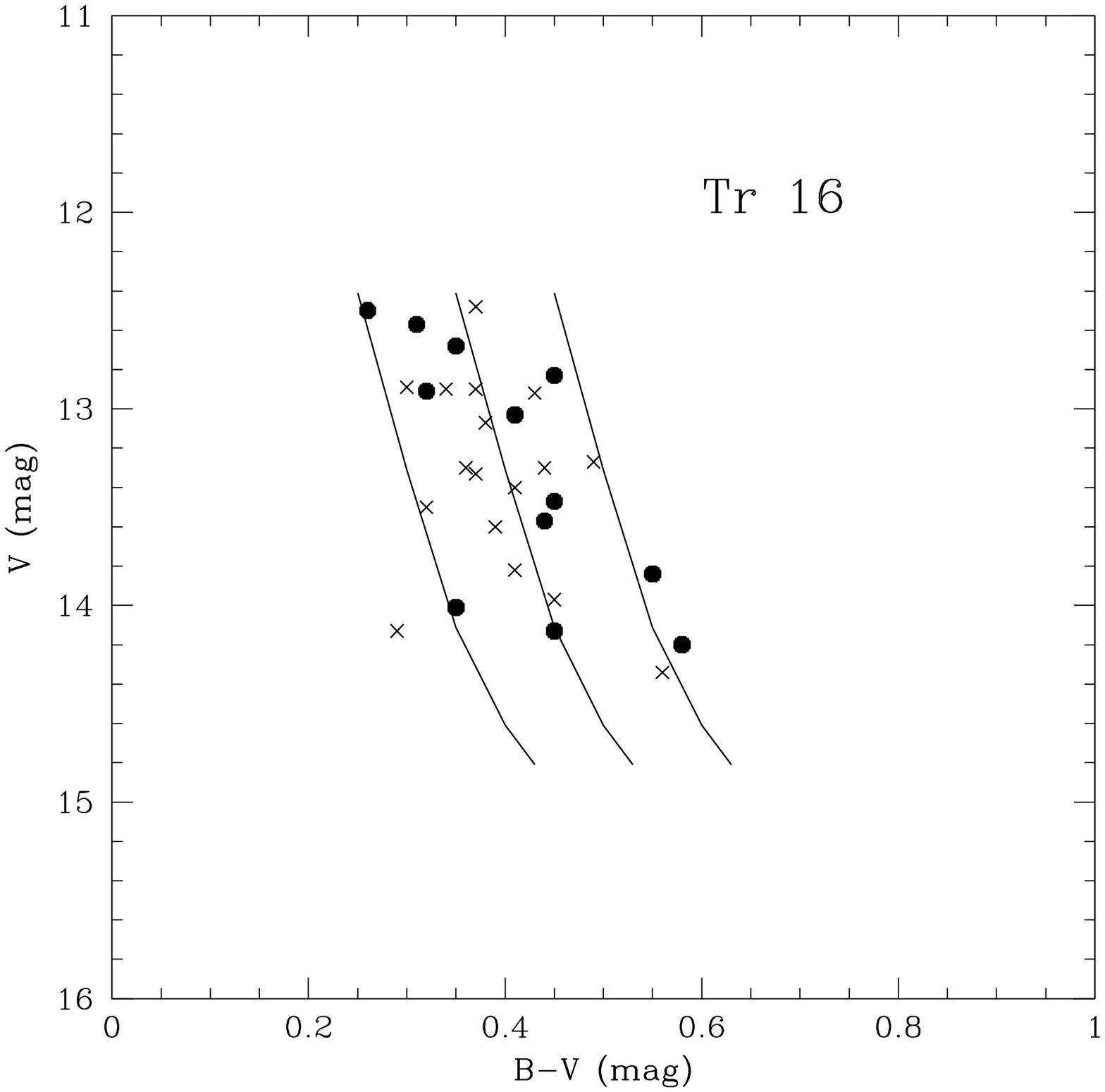}{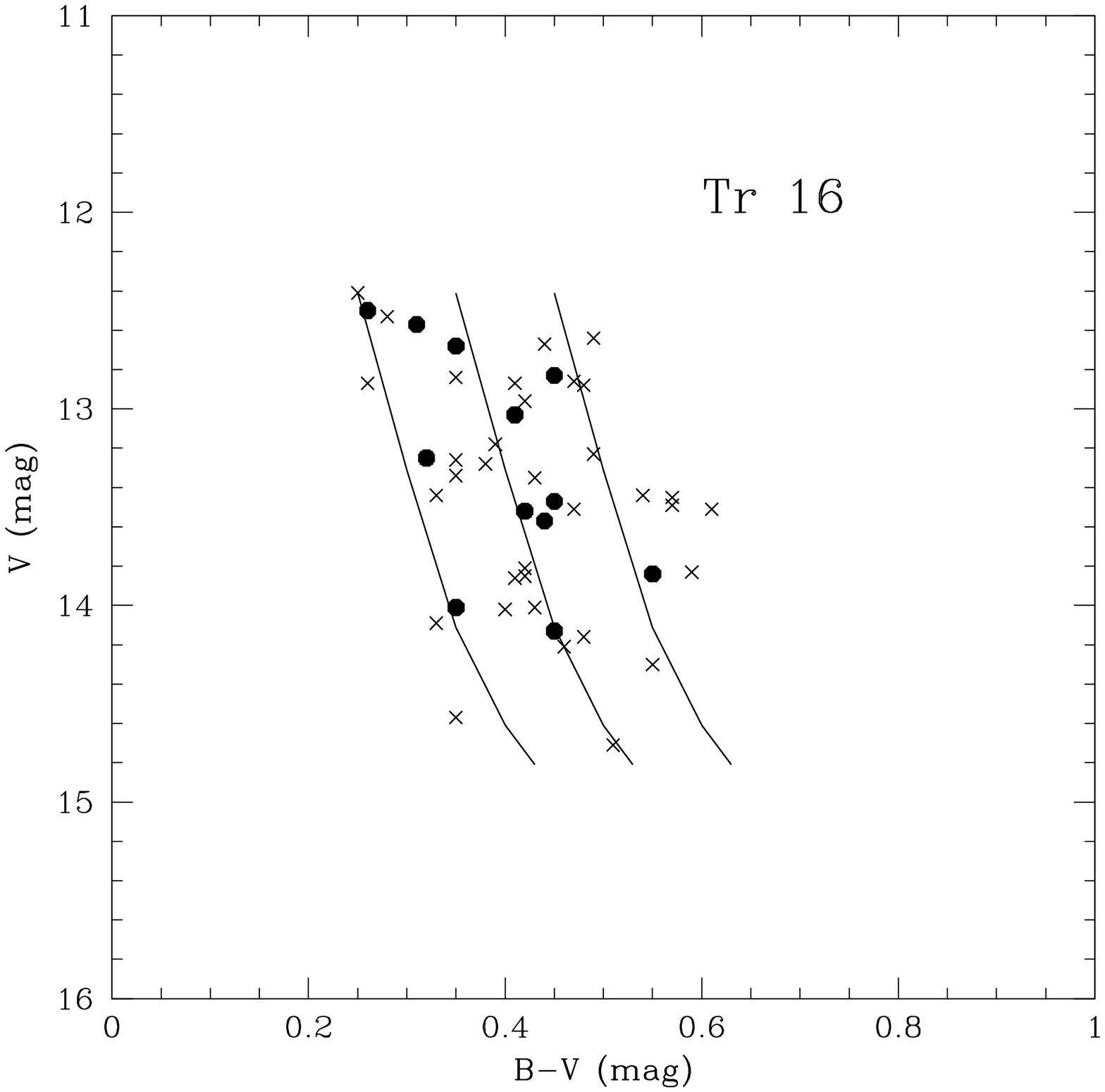}




\plotone{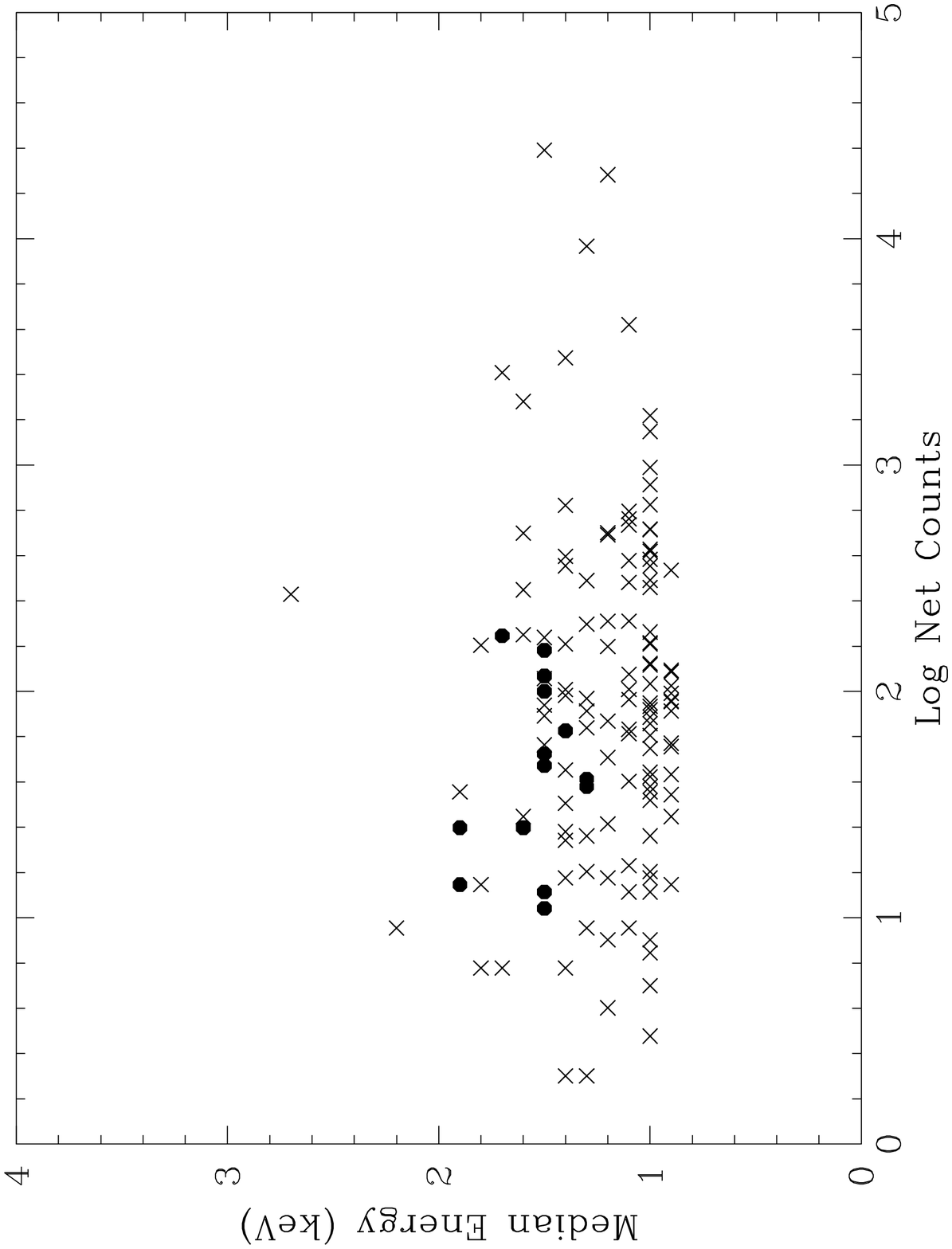}

\plotone{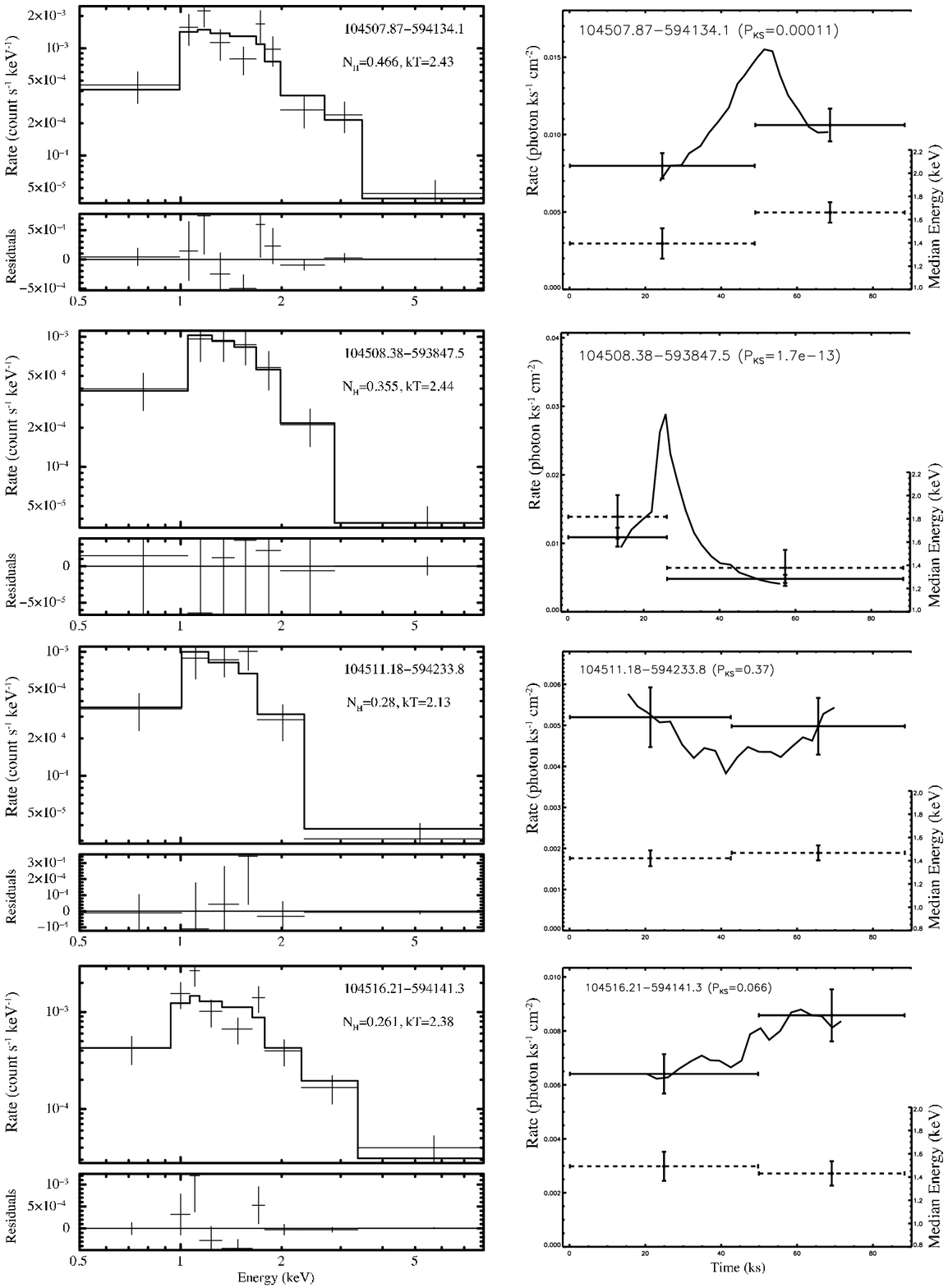}


\begin{thebibliography}{}

\bibitem[albacete etal (2008)]{ac08}Albacete-Colombo, J. F., Damiani, F., Micela, G., Sciortino, S., and
Harnden, F. R., Jr. 2008, A\&A, 490, 1055 (AC) 

\bibitem[babel (1997)]{ba97}
Babel, J. and Montmerle, T. 1997, A\&Ap, 323, 121



\bibitem[broos etal (2010)]{br10}Broos, P. S., Townsley, L. K., Feigelson, E. D., Getman, K. V., Bauer,
F. E., and Garmire, G. P. 2010, ApJ, 714, 1582

\bibitem[Broos et al.(2011a)]{Broos11a} Broos, P.~S., et al.\ 2011, \apjs, submitted (CCCP Catalog Paper)

\bibitem[Caramazza etal (2007)]{ca07}Caramazza, M., Flaccomio, E., Micela, G. Reale, F., Wolk, S. J., and
Feigelson, E. D. 2007, A\&Ap, 471, 645


\bibitem[cudworth etal (1993)]{cu93}Cudworth, K. M., Martin, S. C., and
  DeGioia-Eastwood, K. 1993, \aj, 105, 1822 

\bibitem[degioia-eastwood etal (2001)]{de01}DeGioia-Eastwood, K.,
  Throop, H., Walker, G., and Cudworth, K. M 2001, \apj, 549, 578

\bibitem[drake (1998)]{dr98}
Drake, S. A. 1998, Cont. Obs. Skalnate Pleso, 27, 382 

\bibitem[drake et al (1994)]{dretal94}
Drake, S. A., Linsky, J. L., Schmitt, J. H. M. M., and Rosso, C. 1994,
ApJ, 420, 387

\bibitem[drake etal (2006)]{dretal 3006}
Drake, S. A., Wade, G. A., and Linsky, J. L. 2006, Proc. X-ray
Universe (ESA SP-604), Ed.: A. Wilson, p. 73



\bibitem[evans (1997)]{ev97}  Evans, N. R. 1997, ApJ, 384, 220


\bibitem[evans et al (2003)]{ev03}  Evans, N. R., Seward, F. D., Krauss, M. I., Isobe, T., Nichols, J.,
Schlegel, E. M., and Wolk, S. J. 2003, ApJ, 589, 509

\bibitem[feigelson et al (2011)]{fei11}Feigelson, E. D. et al.  2011, \apjs,
  submitted 

\bibitem[feinstein et al (1973)]{fe73} Feinstein, A., 
Marraco, H. G., and Muzzio, J. C. 1973, \aaps, 12, 331 

\bibitem[Gagne et al.(1997)]{Gagne97}
Gagn\'e, M., Caillault, J.-P., Stauffer, J. R., and Linsky,
J. L. 1997, ApJ, 478, L87


\bibitem[Gagn{\'e} et al.(2011)]{Gagne11} Gagn{\'e}, M., et al.\ 2011, \apjs, submitted (CCCP Massive Star Signatures Paper)

\bibitem[Kouwenhoven et al.(2007)]{Kouwen11}Kouwenhoven, M. B. N.,
  Brown, A. G. A., Portegies Zwart, S. F., and Kaper, L. 2007,
  \aap. 474, 77.

\bibitem[leone (1994)]{le94}
Leone, F. 1994, A\&Ap, 286, 486


\bibitem[mason etal (2009)]{ma09} Mason, B. D., Hartkopf, W. I., Gies, D. R., Henry, T. J., and Helsel,
J. W. 2009, AJ, 137, 3358

\bibitem[massey and johnson (1993)]{mj93}Massey, P. and Johnson, J. 1993, \aj, 105, 980.


\bibitem[Naz{\'e} et al.(2011)]{Naze11} Naz{\'e}, Y., et al.\ 2011, \apjs, submitted (CCCP Massive Star Lx/Lbol Paper)

\bibitem[povich etal (2011)]{pov11}Povich, M. S., et al. \ 2011, \apjs,
  submitted (CCCP IR Excess Paper)

\bibitem[power etal (2007)]{po07}
Power, J., Wade, G. A., Hanes, D. A., Aurier, M., and  Silvester,
J. 2007, in ``The physics of Magnetic Stars", Proc. of Conf. by
Special Astp Obs. of the Russian, AS, eds. I. I. Romanyuk and
D. O. Kudryavtsev, p. 89

\bibitem[preibisch (2005)]{pf06} Preibisch, T. and Feigelson, E. D. 2005, ApJS, 160, 390

\bibitem[schmidt-kaler (1982)]{sk82}
Schmidt-Kaler, T. 1982, in Landolt-B\"ornstein VI 2b, ed K. Schaifers
\& H. H. Voigt (New York: Springer), 19

\bibitem[seward (2000)]{se00}Seward, F. D. 2000, in {\it Allen's
  Astrophysical Quantities}, ed. A. N. Cox, {New York: Springer
  Verlag}, p. 196

\bibitem[schoeller (2010)]{sch10}
Sch\"oller, M., Correia, S., Hubrig, S., and Ageorges, N. 2010, A\&Ap,
522, 85   



\bibitem[smith (2006)]{sm06}Smith, N  2006, \apj, 644, 1151

\bibitem[stelzer etal (2005)]{st05}
Stelzer, B., Flaccomio, E., Montmerle, T., Micela, G., Sciortino,
S., Favata, F., Preibisch, T., and Feigelson, E. D. 2005, ApJS, 160, 557.

\bibitem[Telleschi et al (2007)]{Te07}Telleschi, A., G\"udel, M.,
  Briggs, K. R., Audard, M., and Palla, F. 2007, A\&A, 468, 425


\bibitem[Townsley et al.(2011)]{Townsley11} Townsley, L.~K., et al.\ 2011, \apjs, submitted (CCCP Intro Paper)


\bibitem[Wolk et al.(2011)]{Wolk11} Wolk, S.~J., et al.\ 2011, \apjs, submitted (CCCP Tr16 Paper)


\end{thebibliography}
\end{document}